\documentclass[prd,preprintnumbers,twocolumn,nofootinbib,showpacs,amsmath,amssymb,floatfix]{revtex4}
\usepackage{graphicx}
\usepackage{amsfonts} 
\usepackage{amssymb}
\usepackage{bbm}
\usepackage{epsfig}
\usepackage{graphics}
\usepackage{graphicx}
\usepackage{subfigure}
\usepackage{colordvi}
\usepackage{color}
\usepackage{mathrsfs}
\begin{document}

\def\be{\begin{equation}}
\def\ee{\end{equation}}
\def\bea{\begin{eqnarray}}
\def\eea{\end{eqnarray}}
\newcommand{\rh}{\tilde{\rho}_h}
\newcommand{\dx}{\,\text{d}}
\newcommand{\rs}{\mathscr{R}}
\newcommand{\rsm}{\rs_{\mt{min}}}
\newcommand{\comment}[1]{}
\newcommand{\del}{\partial}
\newcommand{\la}{\lambda}
\newcommand{\mt}{\mathtt}
\newcommand{\dt}{\mathtt{d}}
\newcommand{\al}{\alpha}
\newcommand{\bb}{\beta}
\newcommand{\ga}{\gamma}
\newcommand{\Ga}{\Gamma}
\newcommand{\te}{\theta}
\newcommand{\Te}{\Theta}
\newcommand{\de}{\delta}
\newcommand{\De}{\Delta}
\newcommand{\ka}{\kappa}
\newcommand{\et}{\tilde{e}}
\newcommand{\ze}{\zeta}
\newcommand{\s}{\sigma}
\newcommand{\e}{\epsilon}
\newcommand{\om}{\omega}
\newcommand{\Om}{\Omega}
\newcommand{\La}{\Lambda}
\newcommand{\vp}{\varphi}
\newcommand{\p}{\partial}
\newcommand{\n}{\nabla}
\newcommand{\hn}{\widehat{\nabla}}
\newcommand{\hph}{\widehat{\phi}}
\newcommand{\ah}{\widehat{a}}
\newcommand{\bh}{\widehat{b}}
\newcommand{\ch}{\widehat{c}}
\newcommand{\ddh}{\widehat{d}}
\newcommand{\eh}{\widehat{e}}
\newcommand{\gh}{\widehat{g}}
\newcommand{\ph}{\widehat{p}}
\newcommand{\qh}{\widehat{q}}
\newcommand{\mh}{\widehat{m}}
\newcommand{\nh}{\widehat{n}}
\newcommand{\Dh}{\widehat{D}}

\newcommand{\ti}{\widetilde}
\newcommand{\Ms}{M_{\ast}}
%text
\newcommand{\ie}{{\it i.e.\ }}
\newcommand{\hs}{\hspace{5mm}}
\newcommand{\vs}{\vspace{5mm}\\}

\newcommand{\cE}{{\cal E}}
\newcommand{\cN}{{\cal N}}
\newcommand{\cO}{{\cal O}}
\newcommand{\cR}{{\cal R}}
\newcommand{\cS}{{\cal S}}
\newcommand{\mx}{\mbox}
\newcommand{\ra}{\rightarrow}
\newcommand{\Ra}{\Rightarrow}
\newcommand{\im}{\Longleftrightarrow}
\newcommand{\LF}{\left(}
\newcommand{\RF}{\right)}
\newcommand{\LT}{\left[}
\newcommand{\RT}{\right]}
\newcommand{\hb}{\bar{h}}
\newcommand{\tb}{\bar{t}}
\newcommand{\sqw}{\sqrt{\frac{w}{2}}}
\newcommand{\stwo}{\sqrt{2}}
\newcommand{\2}{\frac{1}{2}}

\newcommand{\mtx}{\mt{max}}
\newcommand{\mtm}{\mt{min}}
\newcommand{\rt}{\tilde{\rho_{\gamma}}}
\newcommand{\rg}{\rho_{\gamma}}
\newcommand{\Vp}{V(\phi)}
\newcommand{\emu}{\e^{2\mu\phi}}
\newcommand{\eal}{\e^{-2\alpha\phi}}
\newcommand{\dphi}{\dot{\phi}}
\newcommand{\ddphi}{\ddot{\phi}}
\newcommand{\ealn}{\e^{2\alpha\phi}}

\date{\today}

\title{Are we seeing the beginnings of Inflation?}

\author{Cosmin Ilie$^1$\email[email: ]{cilie@umich.edu},Tirthabir Biswas$^2$\email[email: ]{tirthabir@yahoo.com} and
 Katherine Freese$^1$\email[email: ]{ktfreese@umich.edu}}

\affiliation{$^1$Michigan Center for Theoretical Physics, Physics Dept., University of Michigan, Ann Arbor, MI 48109, USA\\
$^2$Department of Physics, Pennsylvania State University,
University Park, PA, 16802-6300, USA}
%\affiliation{Physics Dept., University of Michigan, Ann Arbor, MI 48109, USA}

\begin{abstract}
Phantom Cosmology provides  an unique opportunity to ``connect'' the phantom driven (low energy $meV$ scale) dark energy phase to the (high energy GUT scale) inflationary era. This is possible because the energy density increases in phantom cosmology. We present a concrete model where the energy density, but not the scale factor, cycles through phases of standard radiation/matter domination followed by dark energy/inflationary phases, and the pattern repeating itself. An interesting feature of the model is that once we include interactions between the ``phantom fluid'' and ordinary matter, the Big rip singularity is  avoided with the phantom phase naturally giving way to a near exponential inflationary expansion.
\end{abstract}

\maketitle%%%%%%%%%%%%%%%%%%%%%%%
\section{Introduction}
%%%%%%%%%%%%%%%%%%%%%%%
The current accelerated expansion of the universe is usually explained by invoking a Dark Energy (DE)  component\footnote{For alternative approaches which try to avoid dark energy by invoking large scale inhomogeneities see, for instance, \cite{void1,void2,reviews}.}  which today comprises more than $70\%$ of the total energy in the universe (for reviews see \cite{DEreview,Caldwell:2009ix,Silvestri:2009hh}).  The case of a pure cosmological constant, with $w_{\Lambda}\equiv p_{\Lambda}/\rho_{\Lambda}=-1$ marks the divide to the `phantom' realm.  Phantom dark energy models are described  by systems with 
\begin{equation}
w_{p} = \frac{p_p}{\rho_p} <-1
\end{equation}
 and  have the intriguing feature that the energy density in the universe increases with expansion, 
 \begin{equation}
 \rho_p\sim a^{-3(1+w_p)} .
\end{equation}
Hence a universe with  low $\sim mev$ scale accelerated expansion can eventually reach energy scales close to the GUT scale, for instance.  For some examples of cosmological scenarios using phantom energy see \cite{phantomcosmology,bfk}. The question that we want to ask is whether it is possible to exploit this feature of phantom cosmology and turn the dark energy driven acceleration into a GUT scale inflationary phase\footnote{For a brief list of papers that propose various mechanism of connecting the current accelerated expansion to inflation see \cite{acc_infl}}. The idea then would be to construct a cyclic model where dark-energy/inflationary phases are interspersed with decelerating radiation/matter phases.

Several problems immediately appear. Firstly, unless the equation of state for the phantom phase, $w_p$, is extremely close to $-1$, the phantom acceleration will be much faster than the deSitter expansion, and cannot be reconciled with data. Density perturbations produced
during a phantom phase will give rise to a blue spectrum, and consistency with the current WMAP 5-yr data at the $2 \s$ level with tensor modes included~\cite{WMAP5}    requires  $-1>w_{p}>- 1.01$. Secondly, it is well known that phantom cosmology typically ends in a Big-rip singularity, rather than the standard radiation phase which follows inflation. Remarkably, we find that both these problems can be addressed when we include interactions between the ``phantom fluid'' and some hidden sector  matter. Such interactions  ameliorate the phantom acceleration phase to an asymptotic de Sitter type expansion, once the phantom energy density reaches a critical value. It is easy to arrange this transition to occur around the GUT scale, which is appropriate for inflationary cosmology. This also automatically avoids the big rip singularity as the space time now approaches a deSitter universe. The density perturbations
can have a variety of possibilities, allowing for agreement with observations  \cite{WMAP5}.  Moreover, 
in our scenario the universe transitions to an asymptotic deSitter 
phase independent of the value of $\om_p$, and thus avoids having to  fine-tune $w_p$ very close to $-1$. As an additional advantage over the usual slow-roll inflationary scenario, in our phantom-driven inflationary model one does not have to tune the flatness of the potential usually necessary to obtain the large number  of efoldings and near scale-invariant spectrum. In addition, the hierarchy
between  the $\sim meV$ dark energy scale and the GUT inflation scale can re-expressed in terms of four parameters that take values of $\mathcal{O}(1)$ to $\mathcal{O}(10^2)$. We cannot however address the ``coincidence'' problem in our picture.
Finally,  there is the question of how to construct a theoretically self-consistent model of phantom energy. We will comment on this problem
shortly.

 Before delving into the details of our specific realization of the ``phantom cyclic model'', let us 
 outline the  basic picture  by considering just a simple  two fluid model, phantom matter ($\rho_p$) + radiation ($\rho_r$). The cosmology we want to realize is the following:
although the scale factor always increases monotonically with time, the energy density ``cycles'', at least approximately. Each cycle is divided into two different phases: (a) Radiation dominated phase, which starts at an energy density $\rho_r=\la_{\mtx}^4$. As the universe expands radiation gets diluted, the Hubble parameter decreases and reaches a minimum when $\rho_r=\rho_p=\la_{\mtm}^4$. From here on we enter (b) the  phantom energy dominated phase. In realistic cosmology the radiation phase should give way to matter domination at energy densities $\sim (10\ eV)^4$, before giving way to phantom domination, but for simplicity we are going to ignore this slight complication. Thus for a typical scenario which would be consistent with dark energy and inflationary paradigm, $\la_{\mtx}\sim 10^{15}\ GeV\sim 10^{-3}M_p$ or the GUT scale, and $\la_{\mtm}\sim meV\sim 10^{-30}M_p$ corresponding to the scale of current energy density.  Now, in the phantom phase, as the universe expands the energy density increases, and so does the Hubble rate. Initially, depending upon how negative the phantom equation of state parameter, $\om_p$, is this increase in energy density can be quite fast. However, in our model we will see that once the energy density reaches close to a critical scale $\la_{\mtx}$, which is determined by the interactions between the hidden and ordinary matter sector, the energy density and the Hubble parameter asymptote to a constant giving rise to a near exponential expansion. This inflationary phase can end via the reheating mechanism described in section \ref{sec:reheating},~\ref{sec:transition} after which we enter the radiation dominated era of the  next cycle.

A similar idea to the one in this paper has previously been presented by \cite{etexpcyclic}
who dubbed this model "the eternally expanding cyclic universe."  They too (in their Section 4.3)
suggested an alternating increasing/decreasing energy density. However the theory behind
their model is quite different from ours, and consequently their predictions for resultant 
density perturbations are different as well \footnote{They always 
predict a negative, even if extremely small, tilt of the spectral
index while our model will allow a variety of possibilities (as discussed in later sections of this paper).}.
One major problem of all phantom type models is the vacuum stability due to the null energy condition violation.  \cite{etexpcyclic} examine a consistent way to solve this problem, based on a deformation of the ghost condensate model of Arkani-Hamed et all \cite{ghostcondensate}. 

Although the interactions between phantom fluid and ordinary matter can lead to an inflationary 
spacetime, we are still left with a graceful exit problem, or how to ensure that the universe enters the 
standard radiation dominated era. Depending upon the specific model different ``reheating mechanisms'' may be 
able to trigger such a transition. We  focus on a model where the phantom fluid consists of a ghost like 
scalar field coupled to some hidden matter sector. Such a fluid closely resembles the interacting DE-DM 
models~\cite{Amendo,coupled,adinstab,trodden} except that the scalar field instead of being an ordinary  
quintessence field, has negative kinetic energy like a ghost. Although field theory with ghosts is plagued 
with problems of unitarity/instability \cite{ghostinstab},  recent developments attempting
to address these problems
include progress in  non-local~\cite{non-local} and Lee-Wick~ \cite{lee-wick}  higher derivative  models;  
see also \cite{etexpcyclic} and \cite{ghost}.  
As we will see, in our model  the transition from phantom to  radiation phase and vice-versa is achieved partly by suitably choosing the interaction strength between the scalar field and the hidden matter sector, and partly due to the presence of interactions between the hidden and ordinary matter sector. (There is no direct coupling between the ghost field and  ordinary matter.)

As emphasized before, the main reason why the cosmology described above can replace the standard inflationary paradigm is because in the phantom phase the energy density increases even though the universe continues to expand. Thus after the usual dilution in a radiation dominated phase, the phantom phase followed by reheating ensures that the universe again  becomes hot and therefore can reproduce the successes of the Big Bang Model, such as Big Bang Nucleosynthesis and Cosmic Microwave Background Radiation. There is another essential similarity between our scenario and the inflationary paradigm.  The essential reason why inflation solves  the standard cosmological puzzles is because our observable universe (of radius $\sim H_0^{-1}$) can originate from a very tiny region at the beginning of inflation. Something very similar happens in our model as well, the universe expands by a huge factor in every cycle. In our model the number of e-foldings in the radiation and the phantom phase is given by
\be
\cN_{\mt{rad}}\sim \ln\LF \frac{\la_{\mtx}}{\la_{\mtm}}\RF\mx{ and }\cN_{\mt{phan}}\sim \frac{-4}{3(1+\om_p)}\ln\LF \frac{\la_{\mtx}}{\la_{\mtm}}\RF.
\ee
 In order to have a successful GUT scale inflationary paradigm we need $\cN_{\mt{inf}}\gtrsim60$. Thus one gets:
\bea
\cN_{\mt{tot}}&=&\cN_{\mt{phan}}+\cN_{\mt{rad}}+\cN_{\mt{inf}}\\\nonumber
&\sim&\LF1 -\frac{4}{3(1+\om_p)}\RF\ln\LF \frac{\la_{\mtx}}{\la_{\mtm}}\RF+\cN_{\mt{inf}}
\eea
Just to get an idea, if we take $\la_{\mtm}\sim mev$, $\la_{\mtx}\sim 10^{15}Gev$, $\om_p\sim -1.3$, and $\cN_{\mt{inf}}\sim 60$, we get  $\cN_{\mt{tot}}\sim 400$.  What this means is that only a very very tiny portion (for the chosen example, $e^{-400}$ th) of our observable universe will ultimately grow to become the observable universe in the next cycle at the same energy density scale. Another essential similarity between the standard inflationary scenario and our phantom based model, is the production of a huge amount of entropy in every cycle. Even in a ``cyclic'' scenario if one wants to address the usual cosmological puzzles, such as flatness, homogeneity, etc, as well as produce cosmological perturbations with the correct amplitude required for galaxy formation, entropy production seems to be an inevitable requirement~\cite{entropy}. Like in inflation, in our scenario a huge amount of entropy is produced during reheating when most of the phantom energy is converted into radiation. This is essential to ensuring the cyclicity of energy density even though the scale factor of the 
universe is monotonically increasing~\footnote{It may be possible to embed the scenario in phantom cyclic models~\cite{phantom-cyclic,bfk} with actual phases of contraction, but we are not going to explore this possibility here.}.

 It is worth mentioning that in most cyclic models proposed  the scale factor $a(t)$ has a contracting-expanding behaviour, where both a bounce, near the Big Bang, and a turnaround, when the scale factor becomes large are needed. A brief and incomplete compilation of some of the papers that have proposed cyclic cosmologies can be found in \cite{cyclicmodels,bfk}. Such cosmologies provide  natural solutions to the flatness and horizon problems of standard Big Bang scenario. Some variants  also  avoid the issue of initial conditions, provided entropy produced during one cycle is not transferred to the next. In this case  the cycles will not grow, i.e. become larger (Tolman's argument\cite{tolman}) from one to the next, so we can no longer define a beginning of the universe.  For instance the authors of \cite{cyclic} developed  an ekpyrotic inspired cyclical model as an alternative to inflation, where a phase of slow contraction before the bang is responsible for generating a nearly scale invariant spectrum of perturbations that seed the large scale structure formation. In contrast, the standard inflationary scenario assumes a short  phase that occurs after the big bang when the universe is rapidly expanding and nearly scale invariant perturbations are generated.  

The paper is organized as follows. In section \ref{sec:phantom} we present our model of phantom fluid and discuss the cosmology relevant for dark energy. In section \ref{sec:reheating}, we discuss how including interactions can lead to an inflationary space-time along with partially reheating the universe. In section \ref{sec:transition} we provide a specific example where transitions from the phantom-inflation phase to radiation and vice-versa can be orchestrated giving us a cyclic model of the universe. In section \ref{sec:numbers} we discuss the different observational constraints coming from inflation, Big Bang Nucleosynthesis (BBN), and dark energy experiments. Finally, we conclude with a summary of the scenario presented and issues that needs to be addressed further.

\section{Phantom Dark Energy}\label{sec:phantom}
The purpose of this section is to implement a model for the phantom component that would drive the super-accelarated expansion. For now we will not include regular matter nor radiation, but rather focus solely on the components necessary to obtain a phantom phase. One can possibly implement the cosmology sketched before in many different ways. Here we are going  to realize the above picture using a ghost-like scalar field (with negative kinetic energy) $\phi$ coupled to a hidden matter sector which we denote by index ``$h$''. There are two equivalent approaches to describe this type of coupling. One starting from an action, where we allow for a direct coupling term between the hidden sector and the scalar field. The second approach is to consider two fluids that can exchange energy while maintaining the conservation of the total stress-energy tensor, as required by diffeomorphism invariance, although the individual stress-energy tensors are not conserved.

Our system will be described by the following action:
\be\label{actionPDE}
S=\,\int\dx^4x\sqrt{-g}\left[M_p^2 \frac{R}{2}+\frac{(\nabla\phi)^2}{2}+C(\phi)\mathcal{L}_{h}\right]\ ,
\ee
where $\mathcal{L}_h$ does not depend on $\phi$, being the action describing a perfect barotropic fluid. Here we work with a spatially flat FRW metric with signature $(-,+++)$.  Notice that the kinetic term for the scalar field comes with the `wrong' sign, as appropriate for a ghost.
In the phantom dominated phase, described here,  the Hubble equation derived from the action above looks like
\be
H^2= \frac{1}{3M_p^2}(\rho_{\phi}+\rho_{h})=\frac{1}{3M_p^2}\LT-\frac{\dot{\phi}^2}{2}+C(\phi)\tilde{\rho}_{h}\RT\ ,
\label{hubble}
\ee
where we have assumed the field $\phi$ to be homogenous. Here a dot represents the derivative with respect to cosmic time, $t$ and  $\tilde{\rho}_h$ denotes the bare energy density of the hidden sector, which is $\phi$ independent. We will assume that it behaves like a perfect barotropic fluid, satisfying the continuity equation:
\be\label{contbare}
\dot{\tilde{\rho}}_h+3H(\rh+\tilde{p}_h)=0
\ee
with an equation of state
\be\label{defomega}
\tilde{p}_h=\omega\rh
\ee
%For further reference we will use the following notation:
%\be\nonumber
%\rho_h\equiv C(\phi)\tilde{\rho}_h
%\ee
One can also include a potential for the scalar field, and its effects are discussed briefly in the appendix, but for the purpose of illustration we are going to set it to zero.

The interaction that we are going to consider between the hidden matter sector and $\phi$ is going to be very similar to the interactions considered in  coupled quintessence (or interacting DE-DM) models~\cite{coupled, adinstab,trodden}. From the action in equation (\ref{actionPDE}) we get two additional equations of motion.
\bea
\ddot{\phi}+3H\dot{\phi}&=&2\rho_h\mu(\phi)/M_p\label{KG}\\
\dot{\rho}_h+3H(1+\om)\rho_h&=&2\rho_h\dot{\phi}\mu(\phi)/M_p\, ,
\label{cont-h}
\eea
where we have defined
\be\label{Cpara}
\mu(\phi) \equiv\frac{1}{C} \frac{dC}{d\phi}\ ,
\ee
and we are going to assume hence forth that $\mu(\phi)$  always remains positive.
As one can see, both the Klein-Gordon equation for $\phi$ and the continuity equation for the hidden matter sector $\rho_h$ are augmented by interaction terms in the right hand side of the equations (\ref{KG},\ref{cont-h}). Although  $\mu$  in general depends on $\phi$, to understand the phantom phase let us consider a constant $\mu$ to begin with. It is easy to check that the above interaction is consistent with the conservation of the total energy momentum tensor:
\be
\dot{\rho}_{\mt{tot}}+3H(\rho_{\mt{tot}}+p_{\mt{tot}})=0
\ee
where $\rho_{\mt{tot}}\equiv \rho_h+\rho_{\phi}$. To see this we remind the readers that the Klein-Gordon equation can be recast as
$$
\frac{d}{dt}\LT-\frac{\dot{\phi}^2}{2}\RT-3H\dot{\phi}^2=-2\rho_h\mu\dot{\phi}
$$
\be
\Ra \dot{\rho}_{\phi}+3H(\rho_{\phi}+p_{\phi})=-2\rho_{\phi}\dot{\phi}\mu(\phi)\ ,
\label{cont-phi}
\ee
since the energy density and pressure for a phantom scalar field are given by
\be
\rho_{\phi}=p_{\phi}=K=-\frac{\dot{\phi}^2}{2}
\ee
Thus the source terms in the individual conservation equations (\ref{cont-h}) and (\ref{cont-phi}) cancel each other.

One can  solve the hidden matter continuity equation exactly to find
\be
\rho_{h}=\rho_{h0}C(\phi)\LF\frac{a}{a_0}\RF^{-3(1+\om)}
\ee
where we have chosen the convention that at $a=a_0$, $C(\phi)=1$ and $\rho_{h}=\rho_{h0}$. The $a^{-3(1+\om)}$ dependence reflects the usual dilution of the energy density of an ideal fluid with expansion.   Depending upon the scale of energy density relative to the mass of the hidden matter particles, they can either behave as non-relativistic matter $(\om=0)$ or like a relativistic species ($\om=1/3$)\footnote{In this context we note that we have a choice in how we interpret the augmentation of the energy density with growth of $\phi$. One can either think of this growth as simply the increase in mass of the hidden matter particles if the mass depends on $\phi$, or creation of the hidden matter particles through its interactions with $\phi$, or a combination of the two. To keep things simple we are going to assume that the mass of the hidden matter particles remains a constant, but its number density increases.}. We will for most part consider a light degree of freedom, so that approximately it behaves like radiation.

For the special case when $\mu$ is a constant the Coupling function is given by
\be
C(\phi)=e^{2\mu(\phi-\phi_0)/M_p}
\label{exponential}
\ee
Now, coming back to the evolution equations, we only need to solve the Hubble equation (\ref{hubble}) and the Klein-Gordon equation, the latter  simplifying to
\be
\label{KG2}
\ddot{\phi}+3H\dot{\phi}=\frac{2\mu\rho_{h0}}{M_p}e^{2\mu(\phi-\phi_0)/M_p}\LF\frac{a}{a_0}\RF^{-3(1+\om)}\equiv V'_{\mt{eff}}(\phi)
\ee
It is as if the phantom scalar field is evolving under the influence of an ``effective potential'' given by
\be
V_{\mt{eff}}(\phi)=\rho_{h}(\phi)=\rho_{h0}e^{2\mu(\phi-\phi_0)/M_p}\LF\frac{a}{a_0}\RF^{-3(1+\om)}\label{effpotential}
\ee
An important thing to note  is the +ve sign appearing in front of $V'_{\mt{eff}}$ in equation (\ref{KG2}) because $\phi$ is a ghost field with  negative kinetic energy. It is clear now that because of the peculiar properties of the phantom field, $\phi$ actually rolls up the effective potential $e^{2\mu\phi/M_p}$.

We have the following late time attractor power-law solutions:
\bea
a(t)&=&a_0 \LF\frac{t}{t_0}\RF^n\mx{ \& }\phi= \phi_0+pM_p\ln \LF\frac{t}{t_0}\RF\\ \nonumber
&\im&e^{\frac{\phi}{M_p}}= e^{\frac{\phi_0}{M_p}}\LF\frac{t}{t_0}\RF^p
\label{ansatz}
\eea
with
\be
n=-\frac{1-{\omega}}{4\mu^2-3/2(1-\omega^2)}\mx{ and }p=-\frac{4\mu}{4\mu^2-3/2(1-\omega^2)}
\label{power}
\ee
We have verified (see appendix \ref{sec:stability}) that these late time attractors are indeed stable\footnote{We have not investigated whether these solutions suffer from any hydrodynamic instabilities of the nature found in some interacting quintessence models \cite{adinstab}, and we leave this for a future exercise.}.

In this phase the scalar field and the hidden matter are tightly coupled and evolve as a single fluid with an effective equation of state parameter
\be\label{eqstatemat}
\om_{p}\equiv {p_{\phi}+\frac{p_h}{\rho_\phi+\rho_h}}\to-\frac{8}{3}\frac{\mu^2}{(1-\omega)}+\omega
\ee
The asymptotic value is attained during the late time attractor phase. In the phantom phase, expansion of the scale factor is controlled by $\omega_p$, since $a(t)\sim t^{\frac{2}{3(1+\omega_p)}}$. These are analogues to the coupled quintessence solutions~\cite{coupled,adinstab,trodden}. A detailed derivation of all results presented in this section can be found in the appendix \ref{sec:stability}. The crucial thing to note is that  as long as
\be\label{muphantom}
\mu^2>\frac{3}{8}(1-\omega^2)
\ee
we have a phantom phase, i.e. $\omega_p<-1$. In particular for $\om=1/3$ the last condition gives a constraint on $\mu$,
\be
\mu>\frac{1}{\sqrt{3}}
\label{mu-cond}
\ee
In passing we also note that in this phase most of the energy density is actually stored in the hidden sector; one can check that the tracking ratio between scalar field and the hidden matter density is given by
\be
-\frac{\rho_{\phi}}{\rho_{h}}=-\frac{K}{\rho_{h}}=\frac{8\mu^2}{3(1-\omega)^2+8\mu^2}<1
\label{ratio-ph}
\ee

One can also calculate the amount by which $\phi$ evolves during this phase. A straight forward calculation gives us
\be
\De\phi_{p}=\frac{2}{\mu}\LT1-\frac{4}{3(1+\om)}\RT\ln\LF\frac{\la_{\mtx}}{\la_{\mtm}}\RF
\label{phantom-phi}
\ee
Here $\la_{\mtm}$ and $\la_{\mtx}$ represent the energy scale at which the phantom phase begins and  ends respectively. 

In summary, in this section we have found under what conditions a phantom phase could be described by a stable late time attractor for ghost-like fields coupled exponentially  to a perfect fluid. 
The purely phantom sector we have studied in this section is problematic as an inflationary model.
For example, for a constant (purely) phantom equation of state $\om_p<-1$, typically one would obtain a  blue spectrum (see equation (\ref{pps})) which would be inconsistent with the WMAP data. This conclusion is of course valid with the assumption that primordial perturbations are generated mostly during the phantom phase.  Thus unless $\mu$ is fine tuned to be very close to the critical value (which gives rise to $\om_p=-1$), we will not be able to reproduce the inflationary near scale-invariant spectrum.  Thus far we have considered the phantom sector alone; in the next section we include interactions with the standard model which ameliorate some of the problems of a phantom sector alone.

%%%%%%%%%%%%%%%%%%%%%%%%%%%%%%%%%%
\section{``Partial Reheating'' and Late time deSitter Phase}\label{sec:reheating}
In the previous section we realized the phantom phase through an interacting phantom scalar field and hidden matter sector. In the absence of any new physics this phase is going to last till the Big Rip singularity, as is well known in phantom cosmology. In order for the next  cycle to begin we need to first find a ``reheating'' mechanism which converts most of the phantom energy density to radiation. 

What we find, quite remarkably, is that once we include interaction between the hidden matter sector and Standard Model particles (namely we allow the hidden sector particles to be converted to light degrees of freedom of the standard model), which generically exist, it naturally ameliorates the phantom like acceleration to a near exponential inflationary expansion.  

 As we will see, such interactions begin the process of reheating the universe by  producing a radiation bath. Unfortunately the interactions don't provide us with a graceful exit from the inflationary phase, but we will discuss how this issue can also be addressed in the next section.

To understand how interactions effect the cosmological evolution we will use  Boltzmann equations in the following form:
\bea
\dot{\rho}_h+3H(1+\omega)\rho_h&=&-\rho_h\Ga+2\rho_h\dot{\phi}\frac{\mu}{M_p}\label{hidensector}\\
\ddot{\phi}+3H\dot{\phi}&=&2\rho_h\frac{\mu}{M_p}\label{KGannih}\\
\dot{\rho}_{\ga}+4H\rho_{\ga}&=&\rho_h\Ga\label{radiation}
\eea
along with the Hubble equation
\be
H^2=\frac{1}{3M_p^2}\LF\rho_h-\frac{\dot{\phi}^2}{2}+\rho_{\ga}\RF
\ee
Here by $\rho_{\ga}$ we include all the light degrees of freedom which do not couple to the phantom field, and $\Ga$ is the annihilation rate of the hidden matter particles into all these other light degrees of freedom. We have  ignored  the inverse process of creation of the hidden matter particles from the rest of the matter under the assumption that the equilibrium density of the hidden matter sector is small compared to normal radiation. Now, the annihilation rate, per hidden sector particle, is given in general by
\be
\Ga_{h\hb\ra \ga\ga}=n_h<\s|v|>_{h\hb\ra \ga\ga}
\ee
where  $<\s|v|>$ is the average over all initial and final states of the differential cross-section times the relative velocities of the annihilating particles. Usually one is used to consider the opposite process  while trying to determine when a given species freezes out. In the latter case since the photons are in thermal equilibrium, one can use thermal distribution functions to compute the ``thermally averaged''  $<\s|v|>$. However, in our case in order to compute $<\s|v|>$ we would need information regarding the velocity distribution of the produced hidden particles from $\phi$. In the absence of any micro-physical theory of such an interaction, for the purpose of illustration, here we are simply going to assume that $<\s|v|>$ is a constant set by the details of the interaction, so that the interaction rate per hidden sector particle is given by
\be
\Ga=\frac{\rho_h}{m^3}
\ee
Here $m$ is an energy scale we introduce as a free parameter. More generally one expects $\Ga$ to go as some power law, $\Ga\sim \rho_h^{\la}$, the power being determined by the micro-physics, but most of our  results and conclusions should hold qualitatively as long as $\la>1$.

It is easy to check that the above set of equations (\ref{hidensector}-\ref{radiation}) have an asymptotic de Sitter late time attractor solution where all the energy densities and the Hubble parameter tend to a constant. Defining the following dimensionless variables,
\be\nonumber
 \xi\equiv \frac{\rho_h}{M_p^2H^2} \quad\text{ and}\quad \eta\equiv\frac{\Ga}{H}=\frac{\rho_h}{m^3H}
\ee
 we have:
\bea
\xi&\ra& \frac{27\omega-9+3\sqrt{9(3\omega-1)^2+192\mu^2}}{ 8\mu^2}\label{dSx}\\
\eta&\ra& \frac{3\omega-9}{2}+\frac{1}{2}\sqrt{9(3\omega-1)^2+192\mu^2}\label{dSy}\\
\dot{\phi}&=&\frac{2\rho_h\mu}{3H}\ra \frac{2\mu m^3\eta} {3M_p}\label{dSphi}\\
\rho_{\ga}&=&\frac{\rho_h^2}{4H m^3}\ra \frac{m^6\eta^3}{4M^2_p\xi}\label{dSrad}
\eea
As a  consistency check we have evolved these equations  numerically, see Fig.1, and have verified  that the asymptotic values are exactly the ones predicted by the above set of equations.
Let us make a few observations. Since we are specializing to $\om=1/3$, let us look at the asymptotic value of the energy densities in this case:
\bea
\rho_{\ga}&=&\frac{16\,\sqrt{3}\mu\,{m}^{6}}{9M_p^2} \left( \,\sqrt {3}\mu -1\right)^3\nonumber\\
\rho_{h}&=&\frac{16\,\sqrt{3}\mu\,{m}^{6}} {9M_p^2} \left( \,\sqrt {3}\mu -1\right)^2\label{asymptotic}\\
\rho_{\phi}&=&-\frac{32\,\mu^2\,{m}^{6}}{9M_p^2} \left( \,\sqrt {3}\mu -1\right)^2\nonumber
\eea
Note that the solutions above are consistent only if $\mu> \frac{1}{\sqrt{3}}$, but we know this is also a requirement  for the existence of  a phantom phase\footnote{see equation (\ref{mu-cond})} which will eventually settle to this deSitter attractor. In particular, we find that for $\mu\gg 1$, which is the case we will eventually be focusing on, radiation dominates over the hidden sector:
\be
\cR\equiv \frac{\rho_h}{\rho_{\ga}}\approx  \frac{1}{4\sqrt{3}\mu}\ll1\mx{ for }\mu\gg 1
\label{ratio}
\ee
In other words, as $\mu$ increases, the conversion from hidden matter to radiation becomes more efficient, and in particular radiation can easily dominate the total energy density. However, this does not mean that we can enter into the usual radiation dominated epoch because the equation of state for all the energy densities essentially approach $\om_p=-1$, as all the energy densities approach the constant asymptotic values in equation (\ref{asymptotic}). Physically, the energy density that the phantom field was pumping into the hidden matter sector is now transferred to radiation through its interaction with the hidden sector, in such a way that we approach a deSitter universe. This phase however is good for inflationary cosmology as fluctuations produced during this de Sitter phase are expected to give rise to a near scale-invariant spectrum (allowed, although not favored, by  the WMAP data). In order for the amplitude of the fluctuations to be consistent with observations we require
\be
\rho_{\ga}\approx 10^{-12}M_p^4\Ra m\sim 10^{-2}M_p\ .
\ee
We will return to the constraint coming from the number of efoldings required for a successful inflationary paradigm later.

In this section we have found that the big rip singularity could be avoided in this model if the hidden sector particles are converted to light degrees of freedom of the Standard Model. As the energy densities approach their asymptotic values the universe will enter in a deSitter phase. Of the three components, radiation is dominant, yet the universe is inflating. This is due to the interplay between the coupling of the hidden sector to the phantom field and light degrees of freedom of the Standard Model which leads to a state where all the energy densities approach a constant value.   
%%%%%%%%%%%%%%%%%%%%%%%%%%%%%%%%%
\section{Transition to Radiation and Cyclicity}\label{sec:transition}
In the above section we saw that interactions between hidden matter and radiation can ameliorate the phantom like acceleration to exponential inflation but cannot provide a graceful exit from the deSitter inflationary phase. So far we have assumed that $\mu$ remains a constant leading to an exponential coupling between the hidden sector and the phantom field  as in equation (\ref{exponential}). If $\mu(\phi)$ is not a constant a graceful exit becomes possible. Here we explore  the case when $\mu(\phi)$ is periodic. For simplicity we will just assume a step function for $\mu(\phi)$, where
\bea
\mu&=&\mu_p\gg 1 \mx{ for }0<\phi<\phi_R\\
\label{mur}
\mu&=&\mu_r\ll 1 \mx{ for }\phi_R<\phi<\phi_0
\eea
and then the pattern repeats itself. The first phase, when $\mu\gg1$, reproduces the phantom phase discussed  in the previous section, leading eventually to the late time de Sitter like attractor evolution. However, now this inflationary phase ends once $\phi$ reaches the transition value $\phi_R$. We are also going to assume that $\om=1/3$ in this scenario. We remind the readers that $\om$ describes the hidden sector, as in equation (\ref{defomega}). As we will see shortly, the periodicity in $\mu$  will ensure that we enter the standard radiation dominated era which lasts till $\phi$ rolls to $\phi_0$.

The evolution of the universe during one cycle can be described in our model using three phases, as indicated in Fig. 1. Phase I 
provides the  ``Reheating'' from inflationary expansion. Phase II describes a standard radiation dominated era. Then a short phantom phase IIIA ensues, during which the energy is driven from the 
$meV$ to the GUT scale.   This is followed by a deSitter inflationary phase IIIB.  After one such cycle
is complete, the next one begins, again cycling through Phases I, II, IIIA, and IIIB successively.

\subsection{Phase I: Reheating}
\label{reheating}
The reheating process is most easily understood when $\mu_r=0$ (defined in equation (\ref{mur})), so let 
us first focus on this simple case. As we discussed in the previous section, as long as $\mu$ is large, 
although radiation is the dominant energy density, the deSitter phase continues. However, if and when $
\mu$ sharply falls to zero, the phantom phase indeed ends. Two things happen. Firstly, since  initially $
\Ga$ is comparable to $H$, as it can be seen from equation  (\ref{dSy}), there is rapid conversion of the 
hidden matter to radiation, but the hidden matter sector now no longer gets replenished by the scalar
 field.  Secondly, the driving term in the right hand side of the Klein-Gordon equation for the phantom scalar field (\ref{KG}) is now absent and as a result $\phi$ slows due to Hubble damping, 
 $\rho_{\phi}\sim a^{-6}$ and eventually comes to a halt.  At the beginning of the reheating phase  the 
 energy densities of 
radiation, hidden matter and $\phi$ are approximately given by the asymptotic values of  the late time de 
Sitter phase (\ref{asymptotic}). As we had pointed out before, for $\mu\gg1$  radiation is the dominant 
energy density component in this asymptotic phase. The ``reheating'' phase further ensures that 
radiation continues to dominate the energy density.  If one tracks the ratio of the energy densities 
between hidden matter and radiation, $\rs$, then it starts with the asymptotic value given by equation 
(\ref{ratio}), decreases rapidly during conversion, and then approaches a constant, $\rs_{\mt{min}}$, 
once the conversion ends. Note, since $\phi$ slows down, $\rho_h$ redshifts almost as radiation and 
therefore  maintains an approximately constant tracking ratio approaching $\rs_{\mt{min}}$. In appendix 
\ref{sec:ratiomin} we calculated this asymptotic ratio to be
\be
\rsm\approx \frac{1}{6}\frac{\sqrt{3}}{\mu_p^2}\label{rfinalrad0}\mx{ for }\mu_r=0\ ,\mu_p\gg1
\ee

Qualitatively, it turns out that one can distinguish two different regimes in the reheating phase depicted in Fig.1 as phases I.$\mathrm{A}$ and I.$\mathrm{B}$. Numerically we found that even after $\mu\ra 0$, it takes a while for the radiation energy density to start decreasing substantially. The reason is somewhat technical and the reader is referred to the appendix \ref{sec:ratiomin} for details. Intuitively, the main reason is that initially the Hubble damping of radiation is compensated by the hidden matter decays into  radiation,  $4H\rho_{\ga}\sim \rho_h\Gamma$. Since radiation is the dominant component of the energy density, this in turn leads to $H$ being approximately constant, as can be  seen in phase I.$\mathrm{A}$ of Fig.1.
%This can be seen from  (\ref{radiation}), where $\mu$ does not appear directly. Since we start with , the two terms will continue to be at the same order of magnitude, until
Once $\rho_h$ decreases appreciably so that $\rho_h\Gamma\ll 4H\rho_{\ga}$, the radiation energy density starts to decrease appreciably and therefore so does the Hubble rate. This is depicted in phase I.$\mathrm{B}$ of Fig.1. At some point, the conversion from hidden matter to radiation effectively stops, 
marking the end of the reheating phase.   

Although the above discussion has been for $\mu_r=0$,
we note that for a non-zero but small $\mu_r$, the reheating phases (I.$\mathrm{A,B}$) 
follow basically the same pattern as in the $\mu_r=0$ case.
The asymptotic tracking ratio~\footnote{Unlike in the $\mu_r=0$ case where the tracking ratio keeps decreasing and asymptotically approaches $\rsm$, when $\mu_r\neq 0$, $\rsm$ is actually a minimum of the tracking ratio that is attained. Since $\phi$ never comes to a halt, the ratio does increase  from its minimum value  of $\rsm$, but this increase is very slight.} between hidden matter and radiation receives a slight correction:
\be
\rsm\sim \frac{1}{18}\frac{\sqrt{3}(3+8\mu_r)}{\mu_p^2}\label{rfinalrad}\mx{ for }\mu_r\ll1\ ,\mu_p\gg1
\ee
We have checked this numerically, and some of the more technical details  are  discussed in Appendix \ref{sec:ratiomin}.

%%%%%%%%%%%%%%%%%%%%%%%%%%%%%%%%%%%%%%%%%%%
\subsection{Phase II: ``Standard'' Radiation Domination}
After the reheating phase, since hidden matter is no longer converted into radiation, the latter  starts to evolve as $a(t)^{-4}$, and consequently $H\sim 1/2t$ as in the standard radiation dominated era. In the meantime $\phi$ continues to be Hubble damped. Once the scalar field effectively stops evolving, the hidden matter starts redshifting as radiation and thus settles down to its constant tracking ratio given in equation (\ref{rfinalrad0}). In particular, one can see from (\ref{rfinalrad}) that for sufficiently large values of $\mu_p$, this ratio can be quite small and easily satisfy constraints coming from BBN and CMB. BBN/CMB only constrains the abundance of dark radiation component to be less than around 10\%~\cite{bbn}. In Fig.1 we refer to the phase when $\phi$ is being Hubble damped as phase II.$\mathrm{A}$, and the subsequent radiative phase as phase II.$\mathrm{B}$.

If $\mu_r$ is precisely zero, then the radiation  dominated  phase IIA can continue forever because 
$\phi$ will effectively come to a halt, and unless the value of $\phi_0$ is fine-tuned, $\phi$ will never make it to the next large $\mu$ phase. As a result the next phantom phase will not begin and the  cyclic picture cannot be sustained. This is why a small but non-zero value of $\mu_r$ is essential to maintaining cyclicity without having to resort to unnatural fine-tuning.

 For a non-zero but small $\mu_r$, the  reheating phases (I.$\mathrm{A,B}$) and the Hubble damping phase (II.$\mathrm{A}$) follow very much the same pattern as in the $\mu_r=0$ case, as discussed
 at the end of the previous section \ref{reheating}.
The main difference when $\mu_r\neq 0$, as compared to the $\mu_r=0$ case, appears in the radiative phase II.$\mathrm{B}$. This is because the non-zero driving term in the Klein-Gordon equation now ensures that instead of coming to a halt, $\phi$ now tracks radiation.  After the initial phase of Hubble damping, the driving term on the right hand side of the KG equation catches up with the Hubble damping term. At this point the scalar field enters a  phase where its energy density approximately tracks that of radiation. This can be seen in Fig.1 where we have divided the radiation phase into two parts. II.$\mathrm{A}$ refers to the regime when the scalar energy density is still being Hubble damped, while  phase II.$\mathrm{B}$ refers to the tracking phase where both the hidden matter  and the scalar energy densities are tracking that of radiation.

This tracking behavior can be approximately obtained as follows: From the KG equation we have
\be
3H\dot{\phi}=\frac{2\rho_h\mu_r}{M_p}\approx \frac{2\rs_{\mt{min}}\rho_{\ga}\mu_r}{M_p}\approx  6\rs_{\mt{min}}M_p\mu_r H^2
\label{slow-roll}
\ee

In the above we have ignored the $e^{2\mu_r\phi}$ dependence of $\rho_h$, as $\phi$ is rolling very slowly, and $\mu_r\ll1$. We will perform a consistency check later. Also, we have ignored the contributions to the energy density coming from the hidden matter and the scalar field as compared to normal radiation. Again, this is justified as $\cR_{\mt{min}}\ll1$ and $\phi$ is rolling slowly. Choosing the ansatz
\be
\phi=M_pp_r\ln\LF\frac{t}{t_R}\RF\Ra \dot{\phi}=\frac{M_pp_r}{t}
\ee
we find that (\ref{slow-roll}) can indeed be satisfied provided
\be
p_r=\rs_{\mt{min}}\mu_r
\ee
In the above analysis we have used the fact that in a radiation dominated universe $H\sim 1/2t$. In particular our analysis tells us that the tracking ratio between the kinetic energy of $\phi$ and radiation is indeed very small
\be
\frac{K_{\phi}}{\rho_{\ga}}=-\frac{2}{3}(\mu_r \rs_{\mt{min}})^2
\ee
justifying our earlier assumption. In the next subsection we will also see that during this phase $\phi$ evolves rather slowly, so that $C(\phi)$ changes only  by an $\cO(1)$ factor ensuring that the hidden matter indeed  behaves as radiation to a very good approximation.

This  radiation dominated tracking era lasts till $\phi$ reaches $\phi_0$ and rolls over to the large $\mu$ region. The next  phase of phantom domination, phase III.$\mathrm{A}$, can now begin.
%%%%%%%%%%%%%%%%%%%%%%%
\subsection{Cyclicity}
\begin{figure}
\includegraphics[width=.5\textwidth, height=6cm]{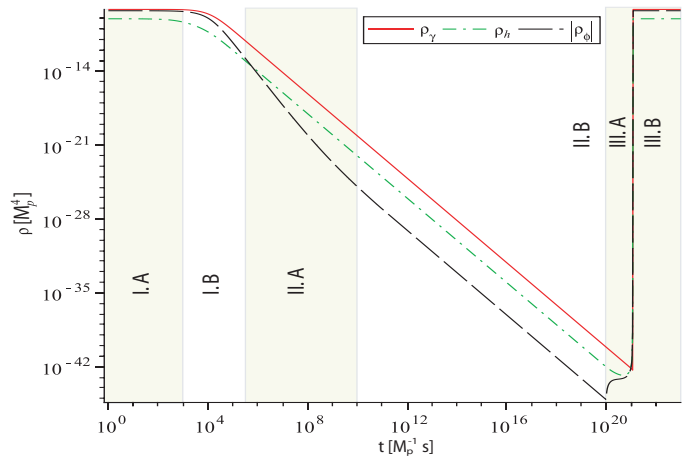}
\caption{Numerical solutions for the energy densities as we complete a cycle from a deStiter phase back to it. Here we have chosen: $\omega=\frac{1}{3}$, $\mu_p=5$, $\mu_r=.1$, $m=10^{-2}M_p$ and we have set $M_p$ to one.  Note the six  distinct phases: I.A and I.B corresponding to reheating; II.A and II.B corresponding to a radiation dominated universe; III.A and III.B corresponding to the phantom and dS phase respectively. In order to make all phases clearly distinct we chose the transition set the minimum energy density at around $10^{-45}$ instead of the realistic $meV^{4}$ }
\end{figure}
To better understand the transition from one cycle to the next let us discuss the various phases we observe in Fig. 1, where we plot a numerical solution for the energy densities of the three components from one deSitter inflationary phase to the next. The plot does not correspond to realistic values for 
$\la_{\mt{max}}$ or $\la_{\mt{min}}$, but captures all the essential features of the different phases. The plot starts (extreme left) at $t=0$  and $\phi=\phi_R$, corresponding to the beginning of the reheating phase I.$\mathrm{A}$.  We have estimated in Appendix  \ref{sec:ratiomin}, equation  (\ref{ttrans}) how long ($t_{1A}$) the I.$\mathrm{A}$ sub-phase lasts
\be
t_{1A}=\left(\frac{3M_p^4\xi}{m^6 \eta^3}\right)^\frac{1}{2}
\ee
During this phase the field $\phi$ evolves approximately a distance of:
\be\label{delta1}
\Delta\phi_{1A}=\dot{\phi}_{dS}\,\int_0^{t_{1A}}\dx t\,e^{-3H_{dS}t}=\frac{\dot{\phi}_{dS}}{3 H_{dS}}\left[1-e^{-3H_{dS}t_{1A}}\right]
\ee
Above we have used:
\be
\dot{\phi}(t_{1A})=\dot{\phi}_{dS}e^{-3H_{dS}t_{1A}}
\ee
where $\dot{\phi}_{dS}$ represents the asymptotic value in the deSitter phase, given by (\ref{dSphi}) and $H_{dS}$ is the Hubble rate during the inflationary phase and can be solved for from the definitions of $\xi$ and $\eta$.

Phase  I.$\mathrm{A}$ is followed by phase I.$\mathrm{B}$, where although the conversion from hidden matter to radiation takes place, the Hubble rate starts to decrease appreciably as well. As soon as the conversion is no longer efficient  we enter a regime, phase II.$\mathrm{A}$, where radiation starts to redshift as $a^{-4}$ marking the beginning of a Standard radiation dominated era. In Phase  II.$\mathrm{A}$ radiation and the hidden sector energies approximately track each other while the field $\phi$ is still being Hubble damped. This phase ends when the scalar field is no longer Hubble damped and starts to track radiation as well, a phase we refer to as II.$\mathrm{B}$. Next we estimate the time $t_{2A}$, when this transition from the Hubble damping  phase II.$\mathrm{A}$ to tracking phase II.$\mathrm{B}$ occurs. It can be defined as the time when the Hubble damping term in the l.h.s. of equation (\ref{KGannih}) is equal to the coupling term on the r.h.s. Under the assumption that we are in a radiation dominated phase and that the hidden sector energy density tracks the radiation energy density with the ratio $\rsm$ we get:
\bea\label{td}
\frac{3\dot{\phi}(t_{2A})} {2 t_{2A}}&=&\frac{2\mu_r}{M_p}\rsm\rho_{\ga}(t_{2A})\nonumber\\
\Ra t_{2A}&=&\frac{t_{1A}^3\dot{\phi}^2(t_{1A})}{M_p^2\mu_r^2\rsm^2}
\eea
This last equation can be rewritten in terms of our parameters in the following form:
\be\label{ratiotttd}
\frac{t_{1A}}{t_{2A}}=\frac{1}{144}\frac{\mu_r^2(3+8\mu_r)^2\eta}{\xi\mu_p^6}e^{6H_{dS}t_{1A}}
\ee
With  $\omega=1/3$ and in the limit $\mu_p\gg 1$  the exponent $6 H_{dS}t_{1A}$ becomes $\sqrt{3}$, so we see, as expected, that $t_{2A}\gg t_{1A}$.

We can compute the distance the field $\phi$ evolves during the phases I.$\mathrm{B}$ and II.$\mathrm{A}$ since the Hubble friction term is dominant during this time. 
\bea\label{delta2}
\Delta\phi_{1B+2A}&=&\int_{t_{1A}}^{t_{2A}}\,\dx t\dot{\phi}=2\dot{\phi}(t_{1A})t_{1A}\left(1-\sqrt{\frac{t_{1A}}{t_{2A}}}\right)\nonumber\\
&\approx&\frac{4\mu_p}{\sqrt{3}}\sqrt{\frac{\xi}{\eta}}M_p e^{-\frac{\sqrt{3}}{2}}
\eea

In the next phase,  II.$\mathrm{B}$, as one can see from the plot,  all the energy components are tracking each other. Since this phase will lasts until almost 'today' and it began at $t_{2A}$ we have:
\be\label{phi-radnonexp}
\frac{\Delta\phi_{2B}}{M_p}\approx 2\rs_{\mt{min}}\mu_r\ln\LF\frac{(\rho_{\ga}(t_{2A}))^\frac{1}{4}} {meV}\RF \approx 120 \rs_{\mt{min}}\mu_r
\ee
where  $\De\phi_{2B}$ is the distance the scalar field evolves during phase II.$\mathrm{B}$. In particular we note that this means
\be
e^\frac{\mu_r\Delta\phi_{2B}}{M_p}\approx e^{120 \rs_{\mt{min}}\mu_r^2}
\label{phi-rad}
\ee
Since $\rs_{\mt{min}}$ is a small number, for sufficiently small values of $\mu_r$ it is easy to see that the exponential will only contribute to an $\cO(1)$ factor to the energy density of the hidden matter sector. In other words unlike the $\mu_r=0$ case, although  the ratio $\rs$ does not monotonically decrease during the radiation phase, but rather starts to increase as $\phi$ evolves, this increase is very slow. This justifies our earlier assumption of hidden matter approximately behaving as radiation.

The advantage of having a non-zero $\mu_r$ is that it keeps the scalar field rolling, albeit slowly, ensuring passage to the next phantom phase when $\phi$ reaches $\phi_0$. Therefore, no fine-tuning is involved in  restarting  the phantom era. To see this observe that
\be
\De\phi_{2B}=(\phi_0-\phi_R)-(\De\phi_{1A}+\De\phi_{1B+2A})
\ee
and crucially for realistic values of the parameters the three different $\De\phi$'s are of the same order of magnitude. If for instance, it turned out that $\De\phi_{2B}\ll\De\phi_{1A}+\De\phi_{1B+2A}$, that would have meant fine tuning the range $\phi_0-\phi_R$ to cancel $\De\phi_{1A}+\De\phi_{1B+2A}$ to very high precision. In fact, this is what one has to do as $\mu_r\ra 0$.

A related nice feature of the model is that the exponential hierarchy between the scales of inflation and dark energy is rather easy to arrange. To see  this let us try to obtain $\la_{\mt{min}}$ in terms of $\phi_0-\phi_R$. By re-arranging equation (\ref{phi-radnonexp}) one finds that the energy density at the beginning of the phantom phase is given by~\footnote{Above we have used a  value for $\rho_{\ga}^{1/4}(t_{2A})$ of $10^{-3}M_P$, slightly overestimating it. The actual value is typically lower than this, because of the additional Hubble dilution of the energy density of radiation from $t_{1A}$ to $t_{2A}$. For instance, with the parameters used in the numerical solution for Fig. 1 we have $\rho_{\ga}^{1/4}(t_{2A})\sim 10^{-5}M_p$.}
$$
\la_{\mt{min}}\sim \la_{\mt{max}}\exp\LF-\frac{\De\phi_{2B}}{2\rs_{\mt{min}}\mu_r M_p}\RF$$
\be\sim 10^{-3}M_p\exp\LF-\frac{\phi_0-\phi_R-\De\phi_{1A}-\De\phi_{1B+2A}}{2\rs_{\mt{min}}\mu_r M_p}\RF\label{hierarchy}
\ee
Since $\rs_{\mt{min}},\mu_r$ are small numbers as compared to $\De\phi/M_p$'s, it is easy to arrange the exponential suppression of $\rho_0$ as compared to the GUT scale. Conversely, we can reformulate the ``smallness'' problem associated with dark energy in terms of four parameters: $\mu_p, \mu_r,\phi_R, \phi_0$, which as we shall see shortly have values that range from $\mathcal{O}(1)$ to $\mathcal{O}(10^2)$.  In  the next section (below equation (\ref{move}))
we will provide specific numerical examples which make this more evident.

Finally, we come to the phantom phases III.$\mathrm{A,B}$. The transition to the phantom phase, which is supposed to be happening during the present cosmological epoch, occurs once $\phi$ reaches $\phi_0$ and $\mu$ transitions suddenly to its high value, $\mu_p$. The hidden matter + scalar field energy densities catch up with radiation, start evolving as a phantom fluid with equation of state given by (\ref{eqstatemat}), and come to dominate the universe. The radiation keeps getting diluted as $a^{-4}$ during this phase which we refer to as phase III.$\mathrm{A}$. Since the increase in energy density in the phantom phase occurs at a very short time scale, all the features of this phase are not very discernable in the log-log plot in Fig.1, but we have checked them numerically. As we have seen before, the annihilation term will ultimately cause this behavior to transition to a de Sitter phase, when $\Ga\sim H$ or $\rho_h\sim m^6/M_p^2$. At this point we enter, what  we  call phase III.$\mathrm{B}$, when all the energy densities become comparable and then tend towards constant values, leading to an asymptotic de Sitter space-time. It is during this phase we expect to generate a scale invariant spectrum of perturbations. This phase will last until the field $\phi$ has evolved a distance of $\phi_0+\phi_R$, when we re-start the cycle.

%%%%%%%%%%%%%%%%%%%%%%%%%%
\section{Numbers and Constraints}\label{sec:numbers}
So far we have provided general constraints coming from different observations on the couplings and the scales. For the purpose of illustration let us provide some typical values which conform to these constraints and in the process we will also be able to understand the different phases of evolution better. Let us start with the inflationary phase. As already discussed, to obtain the correct amplitude of inflationary fluctuations we will take the reheating to occur at approximately GUT scale energy densities which implies
\be
m\sim 10^{-2}M_p
\ee

During the course of 60 e-foldings\footnote{We choose 60 as a generic typical example for GUT scale 
inflation; the actual number of e-foldings might be lower or larger than this.},
we find, using equations (\ref{dSy},\ref{dSphi}), that the field  $\phi$ evolves a distance of:
\be\label{deltainf}
\De\phi_{3B}=40 M_p\mu_p\xi
\ee
As a prototype example, for $\mu_p=4$ this gives us $\De \phi_{3B}\sim 200$. This only gives us a constraint on the range when $\mu$ is large. For instance,  for $\mu_p=4$ the number of efoldings during the phantom phase turns out to be (using equation (\ref{phantom-phi})) $\De\phi_{3A}\sim 30$.  To be consistent with inflation we must have $\phi_R\gtrsim 200+30=230$.  Given the
fact that these field values are transPlanckian, it is possible that higher order terms in the
Lagrangian should not be neglected; we proceed here with the assumption that our starting point
is sensible nonetheless.

Depending on the number of e-foldings during the de Sitter phase we find two distinct cases for the spectrum of fluctuations.
Case I.  If the range of $\phi$ is such that one gets more than 60 e-foldings of de Sitter, 
then the CMB fluctuations are scale invariant. 
As we noted in the introduction, $n_s\approx 1$  is still consistent with observations \cite{WMAP5} once one allows the possibility of running of the tilt and/or tensor modes.  
If the range in $\phi$ is such that we have only around 60 e-foldings then the CMB fluctuations  at large scales can show a transition from a blue (when the phantom phase will be operating and the reheating mechanism hasn't kicked in completely) to a scale invariant spectrum. This could be an interesting and rather unique signature of the model.
 Case II. If the number of efoldings in the dS phase is shorter than 60 e-foldings, the
 fluctuations that we are observing in the sky must have been generated in the phantom phase. This gives a rather stringent constraint on how far below $-1$ the phantom equation of state $\om_p$ can be. For a ``super-inflationary'' space-time sourced by phantom fluid, the spectral tilt is expected to be given by
\be\label{pps}
\eta_s-1=\frac{6(1+\om_p)}{ 1+3\om_p}>0
\ee
implying  a blue spectrum. Therefore to be consistent with the observations $\om_p$ has to be very close to $-1$. For instance, if we include tensor modes, according to \cite{WMAP5} at the $2\s$ level we find
\be
\eta_s<1.01\Ra \om_p>-1.01
\ee
implying $\mu_p<.581$. This bound is very restrictive, remembering that in order to have a phantom phase we need $\mu_p>1/\sqrt{3}\approx 0.577$. In the rest of the section we are not going to discuss this possibility any further and concentrate on Case I with $\mu_p\gg1$ which seems more attractive.

Let's next look at the constraint coming from BBN and WMAP on the amount of dark radiation~\cite{bbn}. To  be consistent with the data the amount of dark radiation has to be limited to within 10\% of ordinary radiation. This essentially imposes a constraint on $\mu_p$, but a rather weak one
\be
\rsm<0.1\Ra \mu_p\gtrsim 0.17
\label{bbn}
\ee
which is easily satisfied.

What about the range of $\Delta\phi$ during radiation domination? This of course depends on the value of $\mu_p,\mu_r$. Just to get an idea, we find for $\mu_p=4,\mu_r=0.2$, we need
\bea
\label{move}
\Delta\phi&=&\De\phi_{1A}+\De\phi_{1B+2A}+\De\phi_{2B}=\phi_0-\phi_R\sim\nonumber\\
&\sim& 0.70+2.16+0.66
\eea
 As one can see, all the values in the above equation 
 are of the same order of magnitude  and therefore no fine tuning seems to be involved.  

Thus the picture that emerges from the above estimates is that for the scenario to work we need a relatively longer phases in $\phi$ when $\mu$ is large (\ref{deltainf}), followed by 
shorter phases when $\mu$ is small.  However, even the discrepancy in $\De\phi$ between
equations (\ref{deltainf}) and (\ref{move}) is only a few orders of magnitude.  Similarly, $\mu_r$
and $\mu_p$ (in the above example) differ again by only a few orders of magnitude.
It is these numbers  (the  values of $\Delta\phi$ and the two values of $\mu$) 
that determine the heirarchy between today's $meV$ scale
and the GUT scale of inflation via equation (\ref{hierarchy}).  So we see that, indeed,
little fine-tuning is required to explain these disparate mass scales.

Finally, let us come to constraints from dark energy. The main constraint comes from the equation of state parameter. A combined $2 \s$ bound from CMB+BAO+SN  is given by $-0.88>\om_p>-1.14$~\cite{WMAP5}. This provides a constraint on $\mu_p$
\be
\mu_p<0.61
\label{mu-constraint}
\ee
This may seem too small, but we realize that currently we are undergoing the phase transition from small $\mu$ to a large $\mu$ region, and therefore it is easy to arrange that the ``current'' value of $\mu$ satisfies inequality (\ref{mu-constraint}) and has not yet reached the constant maximum value $\mu_p$. Equivalently, the equation of state parameter today has not yet reached its late time phantom phase value given by equation (\ref{eqstatemat}).

%%%%%%%%%%%%%%%%%%%%%%%%%%%%
\section{Conclusions}
In this paper we have studied a cyclical model of the universe where the energy density cycles between a minimum value, typically of the order of $meV^{4}$ and a maximum value roughly set by the GUT scale. This effectively provides a connection between the current accelerated expansion we observe today and the inflationary era in the past. The scale factor continues to grow from one cycle to the next
(there is no "turnaround").
In order to achieve this model we postulated the existence of some `hidden' sector matter  coupled  
  to a ghost like scalar field. This mechanism is responsible for a super-accelerated phantom  expansion. Allowing for hidden sector particles to be converted to light degrees of freedom of the standard model ameliorates the phantom behavior, effectively transitioning to a deSitter like expansion, therefore avoiding the Big-Rip singularity. Although dominated by radiation, in this phase all the energy densities remain constant. Therefore, if most of the cosmological perturbations are generated during this exponential inflationary era the spectrum is expected to be scale invariant. Even if not favored by the data, this is still a possibility. Another possibility would be to  have some   of the fluctuations generated during the phantom phase which will show up as a blue tilt in the  spectrum. In that case  we will see a running of the tilt which could be a unique  possible signature of the model. 

In order to achieve the cyclic behavior we have postulated that the coupling between the ghost scalar field and the hidden sector is constant piecewise. This procedure might seem ad-hoc, but it is just the simplest possibility. We found that no fine tuning seems to be involved when requiring to have a long enough radiation/matter dominated phase. It is also worth mentioning that the ``smallnesss'' problem associated with dark energy  is circumvented. The only parameter we need in order to describe the current acceleration is the coupling between the hidden sector and the ghost field, and it has a value not much greater than one. 

\begin{acknowledgments}
T B would like to acknowledge the hospitality of the physics department at University  of Minnesota at Minneapolis. KF and CI are supported by the US Department of Energy and MCTP via the Univ. of Michigan.
\end{acknowledgments}
%%%%%%%%%%%%%%%%%%%%%%%%
\appendix

%%%%%%%%%%%%%%%%%%%%%%%%%%
\appendix
\section{Stability of scaling solutions in phantom coupled models}\label{sec:stability}
We  will look at ghost fields coupled to radiation or matter via the term $\rg=\rt\e^{2\mu\phi}$ and study the stability of the critical points, generalising the analysis done in \cite{Copeland97}. In this way we will find the stable attractors and conditions necessary to obtain them.  Here the subscript $\ga$ refers to the adiabatic index of the fluid the phantom field couples to, defined as $\ga=1+\omega_h$. Throughout the main body of the paper we have used $\rho_h$, but for notational convenience here we will replace it by $\rg$. It will be assumed that the phantom field has a self-interaction potential of the following forms: $\Vp=V_0\eal$ and $\Vp=-V_0\ealn$. We have not used a potential in deriving our main results in the paper, but we keep it here for generality and further reference. 

 For 
 \begin{equation}
 \Vp=V_0e^{-2\alpha\phi} 
 \end{equation}
 we have the following equations:
\bea\label{gov}
\ddot{\phi}+3H\dot{\phi}&=&V_{eff}(\phi)'\nonumber\\
\dot{H}&=&-\frac{1}{2}(-\dot{\phi}^2+\gamma\rt\emu)\\
H^2&=&\frac{1}{3}\left[-\frac{\dphi^2}{2}+\Vp+\rt\emu\right]\nonumber\\\nonumber
\eea
where $V_{eff}=\Vp+\rt\emu$. Although we are treating here only two fluids, this case is relevant during
the phantom phase, when the energy density of any (third) component not coupled to the phantom field, such as regular matter, will quickly become sub-dominant.  Introducing the variables 
\begin{equation}
x=\frac{\dphi}{\sqrt{6}H}
\end{equation}
 and 
 \begin{equation}
 y=\frac{\sqrt{\Vp}}{\sqrt{3}H}
 \end{equation}
  and using $\eta\equiv\log{a(t)}$ as the independent variable, instead of the cosmological time $t$, we can rewrite the equations in the form of an autonomous system supplemented with the Hubble constraint.
\bea\label{autopos}
x'&=&-3x+\sqrt{6}\left[\mu(1+x^2-y^2)-\alpha y^2\right]+\nonumber\\
&&+\frac{3}{2}x\left[\gamma\left(1+x^2-y^2\right)-2x^2\right]\nonumber\\
y'&=&-\sqrt{6}\alpha xy+\frac{3}{2}y\left[\gamma(1+x^2-y^2)-2x^2\right]\\
1&=&\left[y^2-x^2+\tilde{\Omega_{\gamma}}\right]\nonumber\\\nonumber
\eea

In order to find the critical points one has to set the $r.h.s$ of the first two equations to zero and solve for $x$ and $y$. We find five distinct solutions,
\begin{equation*}
(x,y)=\left\{
\begin{array}{lllc}
(i &, & 0)&I.\\
(-i&,  &0)&II.\\
(-\frac{2}{3}\frac{\sqrt{6}\mu}{-2+\gamma}&, & 0)&III.\\
(\frac{1}{4}\frac{\sqrt{6}\gamma}{\alpha+\mu}&, & \frac{1}{4}\frac{\sqrt{2}\sqrt{8\mu\alpha+8\mu^2-6\gamma+3\gamma^2}}{\alpha+\mu})&IV.\\
(-\frac{1}{3}\sqrt{6}\alpha &, & -\frac{1}{3}\sqrt{3}\sqrt{3+2\alpha^2})&V.
\end{array}\right.
\end{equation*}
Note that the first two are nonphysical so we will no longer consider them. Next we compute the fractional densities for the 'phantom' field and its adiabatic constant near the five critical points. The Hubble constraint will then enforce additional existence conditions.
\be\label{relic}
\Omega_\phi=y^2-x^2=\left\{
\begin{array}{lc}
-\frac{8}{3}\frac{\mu^2}{(-2+\gamma)^2}&III.\\
-\frac{1}{4}\frac{-4\mu\alpha-4\mu^2+3\gamma}{(\alpha+\mu)^2}&IV.\\
1&V.
\end{array}\right.
\ee
 One solution is completely dominated by the scalar field $\phi$ and the other two exhibit a scaling behaviour. For the first of those the Hubble constraint will not give any additional inequalities in the parameter space since $\Omega_{\phi(III)} $ is clearly negative, hence less than 1. Here the roman numeral subscript refers to all the solutions, including the non-physical ones, i.e $(I)$ corresponds to the solutions with $(x,y)=(i,0)$ and so on.  Also $\Omega_{\phi(IV)}<1$ is trivially satisfied for positive parameters. So the only nontrivial existence constraint comes from imposing reality of the $y_{IV}$ solution:
\be\label{existence}
\mu(\mu+\alpha)\geq\frac{3}{8}\gamma(2-\gamma)
\ee
 Let us now look at the 'effective' values of the adiabatic constant $\gamma=1+w$ for the ghost field near the critical points. For completeness we will keep even the nonphysical solutions.
\be
\ga_{\phi}=1+\frac{p_{\phi}}{\rho_{\phi}}=\frac{\dot{\phi}^2}{\2\dphi^2-\Vp}=\frac{2x^2}{ x^2+y^2}
\ee
With this definition we get,
\be\label{adconstant}
\gamma_\phi=\frac{2x^2}{-y^2+x^2}=\left\{
\begin{array}{lc}
2&I.\\
2&II.\\
2&III.\\
-\frac{3\gamma^2}{-4\mu\alpha-4\mu^2+3\gamma}&IV.\\
-\frac{4}{3}\alpha^2&V.
\end{array}\right.
\ee
The total, or 'effective' DE equation of state parameter that drives the expansion can be defined since we have scaling solutions.
\be
\omega_{tot}=\frac{p_{\phi}+p_{\gamma}}{\rho_{\phi}+\rho_{\ga}}
\ee
Some algebra leads to
\be\label{wtotgeneral}
\omega_{tot}=\Omega_{\phi}(\ga_{\phi}-1)+(1-\Omega_{\phi})(\ga-1)
\ee
For the three physical solutions it leads to the following values:

\be\label{wtotalpospot}
\omega_{tot}=\left\{
\begin{array}{lc}
\frac{1}{3}\frac{3\omega_h(\omega_h-1)+8\mu^2} {\omega_h-1}&III.\\
\frac{1}{2}\frac{3(1+\omega_h)^2+2(\mu+\alpha)(\omega_h\alpha-\mu)}{(\alpha+\mu)^2}&IV.\\
-1-\frac{4}{3}\alpha^2&V.
\end{array}\right.
\ee
Next we  will study the stability of the  relevant critical points, ignoring the non-physical first two solutions. The technique is the following. One expands around the critical solution setting $x=x_c+u$ and $y=y_c+v$ into (\ref{autopos}) and then keep only linear terms. In order to have a stable solution the eigenvalues of the matrix describing the linearized system must have negative real parts.
For solution $\mathrm{III}$.  one finds the following equations
\bea\label{line3}
u'&=&-\frac{1}{2}\frac{-8\mu^2-3\gamma^2+12\gamma-12}{-2+\gamma}u\nonumber\\
v'&=&-\frac{1}{2}\frac{(-3\gamma^2-8\mu\alpha-8\mu^2+6\gamma)}{-2+\gamma}v
\eea
For both radiation ($\gamma=\frac{4}{3}$) and dust ($\gamma=1$) the coefficient $u$ is negative independent of the value of $\mu$. From the second equation one gets that the node is stable when:
\be\label{stab3}
3\gamma(2-\gamma)<8\mu(\alpha+\mu)
\ee
Otherwise we have a saddle point. It is interesting to notice that this is exactly the same condition we got in (\ref{existence}) for the existence of the fourth solution.  

Let us move on to the stability of solution $\mathrm{IV}$, for which  the linearized system becomes:
\begin{widetext}
\bea\label{line4}
u'&=&-\frac{1}{8}\frac{24\al^2-24\ga\mu^2+24\mu^2+48\mu\al+18\ga^2-36\ga\al\mu-9\ga^3-12\ga\al^2}{(\mu+\alpha)^2}u\nonumber\\
&&-\frac{1}{8}\frac{\sqrt{3}\sqrt{8\mu\al+8\mu^2-6\ga+3\ga^2}(16\mu\al+3\ga^2+8\mu^2+8\al^2)}{(\al+\mu)^2}v \nonumber\\
v'&=&\frac{1}{8}\frac{\sqrt{3}\sqrt{8\mu\al+8\mu^2-6\ga+3\ga^2}(-4\al^2+3\ga^2-4\mu\al-6\ga)}{(\al+\mu)^2}u-\frac{3}{8}\frac{\ga(8\mu\al+8\mu^2-6\ga+3\ga^2)}{(\al+\mu)^2}v
\eea
\end{widetext}
The eigenvalues are:
\bea\label{ev4}
e_{1(4)}&=&\frac{1}{4}\frac{1}{\al+\mu}\left(3\al\ga-6\al-6\mu+B^{\frac{1}{2}}\right)\nonumber\\
e_{2(4)}&=&\frac{1}{4}\frac{1}{\al+\mu}\left(3\al\ga-6\al-6\mu-B^{\frac{1}{2}}\right)
\eea
where $B$  is the following combination of the parameters:
\be\label{B}\nonumber
\begin{array}{lll}
B&=&72\mu\al\left(\ga^2+1+\frac{8}{3}(\mu+\al)^2\right)+54\ga^2(\ga-2)+\\
&&+\al^2(36+81\ga^2-180\ga)+36\ga(\ga-\al\mu+4\mu^2)
\end{array}
\ee
The study of stability of the solutions is quite complicated here but there is a range of parameters for which the real part of the two eigenvalues is negative. Independent of the value of $B$ one necessary condition for stability here is
\be\label{stab4}
2(1+\frac{\mu}{\al})<\ga
\ee
For positive parameters, and if $\ga$ is either $1$ or $\frac{4}{3}$ this inequality cannot be satisfied;
thus the fourth attractor is unstable if the hidden sector is comprised of some component that behaves 
like matter or radiation. 

Let us look now at the stability conditions for the critical point labeled by $\mathrm{V}$. The linearized autonomous system becomes in this case:
\bea\label{line5}
u' &=&(-3+2\al^2(\ga-3)-4\mu\al)u+\nonumber\\
&&+\sqrt{2}\sqrt{3+2\al^2}(2\mu-\al\ga+2\al)v\\
v' &=&\sqrt{2}\sqrt{3+2\al^2}(\al\ga-\al)u-\ga(3-2\al^2)v\nonumber
\eea
The eigenvalues read $-(3+2\al^2)$ and $-(4\al^2+4\mu\al+3\ga)$ which are clearly both negative and real so one has a stable node as a late time attractor. Also remember that this solution does not have a scaling behavior, since in this case the energy density is dominated by the phantom. In conclusion we found two stable late time attractors, one of which exhibits a scaling behaviour. 
Notice also that this scaling solution corresponds to a critical point  where $y=0$, so in effect it is equivalent to the case where the potential is actually zero.

Next let us look at the tracking ratio in the two cases of interest, namely the third and fourth critical points. The tracking ratio is defined to be the ratio of relic densities of the two components.
\be\label{trac}
r=\frac{\Omega_{\phi}}{1-\Omega_{\phi}}
\ee
With this definition we find, using (\ref{relic}):
\begin{equation*}
r=\left\{
\begin{array}{rc}
-\frac{8}{3}\frac{\mu^2}{(-2+\ga)^2\left(1+\frac{8}{3}\frac{\mu^2}{(-2+\ga)^2}\right)} & III.\\
-\frac{1}{4}\frac{-4\mu\al-4\mu^2+3\ga}{(\mu+\al)^2\left(1+\frac{1}{4}\frac{-4\mu\al-4\mu^2+3\ga}{(\mu+\al)^2}\right)} & IV.
\end{array} \right.
\end{equation*}

Since we have established that we will be using the late time attractor described by critical point $\mathrm{III.}$ as our model for the phantom phase, let us actually see under what conditions we cross the 'phantom divide' and if this phase will be stable. Comparing $\omega_{tot}$ given by (\ref{wtotalpospot}) the following restriction on $\mu$:
\be\label{accmat}
\mu^2\ge\left\{
\begin{array}{rl}
\frac{3}{8} & \text{for matter hidden sector}\\
\frac{1}{3} & \text{for radiation hidden sector}
\end{array}\right.
\ee
Looking back at (\ref{stab3}) and taking $\alpha=0$ as appropriate for the case of no potential one can check that the stability conditions are actually identical with the conditions for achieving a phantom phase, listed above. Hence, we have verified that the late time attractor solution for $V(\phi)=0$ is stable once we have $\mu$ such that we get a phantom phase.

Now for generality we will  repeat the same analysis for  a negative potential,
\begin{equation}
V(\phi)=-V_0\exp{2\al\phi}. 
\end{equation}
Here we will define $y=\frac{\sqrt{-\Vp}}{\sqrt{3}H}$.In this case the autonomous system will take the following form
\be\label{auton}
\begin{array}{lcl}
x'&=&-3x+\sqrt{6}\left[\mu(1+x^2+y^2)-\alpha y^2\right]+\\
&&+\frac{3}{2}x\left[\gamma\left(1+x^2+y^2\right)-2x^2\right]\\
y'&=&\sqrt{6}\alpha xy+\frac{3}{2}y\left[\gamma(1+x^2+y^2)-2x^2\right]\\
1&=&\left[-y^2-x^2+\tilde{\Omega_{\gamma}}\right]
\end{array}
\ee
There will be only two physical critical points in this case:
\begin{equation*}
(x,y)=\left\{
\begin{array}{lllc}
(-\frac{2}{3}\frac{\sqrt{6}\mu}{-2+\gamma}&, & 0)&I.B\\
(\frac{1}{4}\frac{\sqrt{6}\gamma}{-\alpha+\mu}&, & \frac{1}{4}\frac{\sqrt{2}\sqrt{8\mu\alpha-8\mu^2+6\gamma-3\gamma^2}}{-\alpha+\mu})&II.B\\
\end{array}\right.
\end{equation*}
Just to clear any possible confusion please note that the labels $I.B$ and $II.B$ here are not related to the labeling of the different regions in Fig. 1. Notice that the first critical point in this case is identical to the third critical point for the case of positive potential. (see equation below (\ref{autopos})) The existence condition for the second critical point reads $\mu(\mu-\alpha)<\frac{3}{8}\gamma(2-\gamma)$.
Fractional density values are
\be\label{relicn}
\Omega_\phi=-y^2-x^2=\left\{
\begin{array}{lc}

-\frac{8}{3}\frac{\mu^2}{(-2+\gamma)^2}&I.B\\
\frac{1}{4}\frac{-4\mu\alpha+4\mu^2-3\gamma}{(-\alpha+\mu)^2}&II.B\\
\end{array}\right.
\ee
From here we get an additional existence constraint on the second critical point, namely:
\be\label{constrain2neg}
\alpha(\mu-\alpha)\leq\frac{3}{4}\gamma
\ee
The adiabatic constant for $\phi$ is
\be\label{adconstantn}
\gamma_\phi=\frac{2x^2}{y^2+x^2}=\left\{
\begin{array}{lc}

2&I.B\\
-\frac{3\gamma^2}{-4\mu\alpha+4\mu^2-3\gamma}&II.B\\
\end{array}\right.
\ee

Notice the difference between the last value in (\ref{adconstantn}) and the fourth in (\ref{adconstant}). Here we can set $\mu\to0$ and recover $\gamma$ as the adiabatic index. Before the naive limit was $-\gamma$ but one was not actually allowed to take that limit due to (\ref{existence}). Now the inequality has been reversed so  the second critical point in the case of the negative potential could be used for  a scaling solution in a phase where the hidden sector decouples from the ghost field.

As before let us look at the effective equation of state parameter:
\be\label{wtotnegpot}
\omega_{tot}=\left\{
\begin{array}{lc}
\frac{1}{3}\frac{3\omega_h(\omega_h-1)+8\mu^2}{\omega_h-1}&I.B\\
\frac{\omega_h\alpha+\mu}{\alpha-\mu}&II.B
\end{array}\right.
\ee

Expanding around the critical points for the first solution we get the following conditions for stability:
\bea\label{stabn1}
0&>&-\frac{1}{2}\frac{-8\mu^2-3\gamma^2+12\gamma-12}{-2+\gamma}\nonumber\\
0&>&-\frac{1}{2}\frac{(-3\gamma^2+8\mu\alpha-8\mu^2+6\gamma)}{-2+\gamma}
\eea

The first inequality is trivially satisfied and the second implies:
\bea\label{stabn11}
\gamma(2-\gamma)<\frac{8}{3}\mu(\mu-\alpha)
\eea
This could  be used as a constraint on the steepness of the potential in order to preserve the tracking behavior in the phantom phase. For the second critical point we have the following eigenvalues:
\bea\label{ev2c}
e_{1(2)}&=&\frac{1}{4}\frac{1}{\al-\mu}\left(3\al\ga-6\al+6\mu+B_2^{\frac{1}{2}}\right)\nonumber\\
e_{2(2)}&=&\frac{1}{4}\frac{1}{\al-\mu}\left(3\al\ga-6\al+6\mu-B_2^{\frac{1}{2}}\right)
\eea
Here $B_2$ is a combination of the parameters, the exact form of which we will not be using. A necessary condition for the stability of this critical point is:
\be\label{stab2neg}
\alpha(1-\omega_h)<2\mu
\ee

The tracking ratios can be easily computed:
\begin{equation*}
r=\left\{
\begin{array}{rc}
-\frac{8}{3}\frac{\mu^2}{(-2+\ga)^2\left(1+\frac{8}{3}\frac{\mu^2}{(-2+\ga)^2}\right)} & I.B\\
-\frac{-4\mu\al+4\mu^2-3\ga}{-4\al^2+4\al\mu-3\ga} & II.B
\end{array} \right.
\end{equation*}

In this appendix  we have found the conditions for scaling behavior in the phantom phase, 
for the cases of positive, zero, and negative exponential potentials. 
An exponential coupling between a ghost like field and some hidden sector fluid is used to generate a 
phantom phase, where the effective equation of state is less than negative one.  
We found the stable attractors and conditions necessary to obtain them.
As a side result we notice that for negative potentials one could have a scaling stable  late time attractor even if the ghost field is decoupled from the hidden sector. In that case the ghost  will just track the hidden sector component.
%%%%%%%%%%%%%%%%%%%%%%%%%%%%%%%%%%%%%%%%%%%%%%%%%%%%%%%%%
\section{Estimating the minimum value of $\rs$}\label{sec:ratiomin}

Here we will estimate the minimum value of the ratio between the hidden sector energy density and the radiation energy density 
\begin{equation}
\rs\equiv\frac{\rho_h}{\rho_{\ga}} . 
\end{equation}
during the reheating phase (see Fig.1 regions $\mathrm{I.A}$ and $\mathrm{I.B}$). Once we enter the transition phase (region $\mathrm{I.A}$), if $\mu$ is large enough, most of the energy density will be stored in radiation, followed by the phantom field, and the least amount is contained by the hidden matter sector. To a good approximation we will use 'radiation domination' in what follows, since we are only interested in order of magnitude estimates.

First we will look at the case when the hidden matter is non-relativistic when this transition occurs, i.e.  $\omega=0$. Setting $\mu_r$ to zero, as appropriate for the transition to radiation, we will have the following equations for the evolution of the system:
\bea
\dot{\rho}_h+3H\rho_h&=&-\rho_h\Ga\\
\ddot{\phi}+3H\dot{\phi}&=&0\label{KGrmin}\\
\dot{\rho}_{\ga}+4H\rho_{\ga}&=&0\label{aproxradcont}
\eea
along with the Hubble equation
\be
H^2=\frac{1}{3M_p^2}\LF\rho_{\ga}\RF
\ee
We have neglected the annihilation term in equation (\ref{aproxradcont}) because $\rho_h\Ga\ll 4H\rho_{\ga}$ will be satisfied very quickly after the transition from the deSitter phase is started. The solutions for the energy densities are simple:
\bea
\rho_{h}(t)&=&\frac{1}{t^{\frac{3}{2}}C_1-2t m^{-3}}\label{hiddentrans}\\
\rho_{\ga}(t)&=&\frac{3M_p^2}{4t^2}\label{radtrans}
\eea
where $C_1$ is an integration constant we still have to fix.  We have used the fact that during radiation domination $H\sim \frac{1}{2t}$ in order to get the coefficient for the radiation energy density. Since it takes a longer time for the annihilation term to be sub-dominant with respect to the Hubble dampening term in the hidden matter continuity equation, we will have a short period where the energy density in hidden matter is decreasing faster than $a^{-3}$.  This yields a minimum value for the parameter $\rs$. 
Simply setting $\dot{\rs(t)}=0$ leads to $\Ga(t_c)\sim H(t_c)$. In order to fix the integration constant 
$C_1$, we need to know at what time we start the transition from deSitter to radiation dominated phase. Using the initial condition\footnote{Here we consider the transition from dS phase to the radiation dominated phase due to the sudden drop in the coupling between the hidden sector fields and the ghost field. The asymptotic value (valid for the dS phase) for the energy density in equation (\ref{dSrad})  (the right hand side of the equation) here becomes an initial condition for the transition phase, Region IA.}  
in (\ref{dSrad}) along with equation (\ref{radtrans}), the transition time is found to be:
\be
t_{1A}=\left(\frac{3M_p^4\xi}{m^6 \eta^3}\right)^\frac{1}{2}\label{ttrans} .
\ee

Furthermore, $C_1$ is obtained by requiring that the hidden matter density $\rho_h$ value in the deSitter phase matches the value estimated at the transition time $t_{1A}$, obtained from (\ref{hiddentrans}). The initial conditions (\ref{dSy}) together with (\ref{dSx}) allows us to evaluate the initial value of hidden matter density at the beginning of the transition phase. In the limit of $\mu_p \gg 1$ we get the following simplified form for the integration constant:
\be
C_1=\frac{\sqrt{3}(3 M_p^2+32\sqrt{3}\mu_p^3 m^3 t_{1A})}{48 m^6 \mu_p^3 t_{1A}^\frac{3}{2}}\label{intconst}
\ee
In order to get the minimum value for $\rs$ we go back to the condition $\Ga=\frac{\rho_h(t_c)}{m^3}\sim H(t_c)$, where $t_c$ represents the time at which this minimum is attained. This can be rewritten as:
\be
\frac{\rho_h}{m^3}\sim \frac{\rho_{\ga}^{\frac{1}{2}}}{\sqrt{3}M_p}\label{critcondition}
\ee
leading to 
\be
\rsm\sim \frac{2 m^3 t_c }{3 M_p^2}\label{rcrit}
\ee
Also, from (\ref{critcondition}) we can solve for $t_c$ using (\ref{hiddentrans}) and (\ref{radtrans}): $t_c=\frac{16}{C_1^2m^6}$. Plugging into (\ref{rcrit}) we get the simplified form:
\be
\rsm=\frac{32}{3 M_p^2 m^3 C_1^2}
\ee
Using (\ref{ttrans}) and (\ref{intconst}) in the above equation we obtain the final result, expressed only in terms of $\mu_p$:

\be
\rsm=\frac{16}{\sqrt{3}\mu_p^\frac{3}{2}(1+4\sqrt{\mu_p})^2}\label{rfinal}
\ee

Let us now turn our attention to the case where the hidden matter becomes relativistic at the energy scales where the transition between the deSitter and the radiation phases occurs. At the beginning there will still be a regime where the annihilation is effective, thus lowering the value of the ratio between the hidden matter energy density and the radiation energy density. If $\mu_r=0$, instead of a minimum we will now have a constant asymptotic value towards which this ratio will tend. This is due to the fact that  once the conversion is no longer efficient, both hidden matter and radiation energy densities will scale as $a^{-4}$. For completeness and generality we will study the case where $\mu_r\neq 0$. As we shall shortly see here a minimum develops, just as we have seen in the case of non-relativistic hidden matter. The approximative equations we need to solve are:
\bea
\dot{\rho}_h+4H\rho_h&=&-\rho_h\Ga+2\rho_h\dot{\phi}\frac{\mu_r}{M_p}\label{hidradhm}\\\
\ddot{\phi}+3H\dot{\phi}&=&0\label{KGradhm}\\
\dot{\rho}_{\ga}+4H\rho_{\ga}&=&0\label{radradhm}
\eea
along with the Hubble equation
\be
H^2=\frac{1}{3M_p^2}\rho_{\ga}
\ee
Notice that we have neglected the $\rho_h$ terms in the second and third equations since we are interested in the phase of rapid conversion of hidden sector particles to radiation. Therefore, in this regime the energy density stored in the hidden sector will decay much faster than $a^{-4}$. Essentially the reason for this being that $\mu$ has transitioned from large to small values, i.e $\mu_p\gg\mu_r$. The system admits the following solutions:

\bea
H(t)&=&\frac{1}{2t}\\
\rho_{\ga}(t)&=&\frac{3M_p^2} {4t^2}\label{radradtrans}\\
\dot{\phi}(t)&=&\frac{A_{\phi}}{t^\frac{3}{2}}\label{pradtrans}\\
\rho_{{h}} \left( t \right)& =&\frac{8\,{m}^{3}{A_{{\phi}}}^{2}{\mu _{{r}}}^{2}}{D(t)}\label{hradtrans}
\eea
Where D(t) is:
\be
 D(t)= 4\,{t}^{3/2}M_{{p}}A_{{\phi}}\mu _{{r}}+{t}^{2}{M_{{p}}}^{2}+8\,{t}^{2}{{\rm e}^{4\,{\frac {A_{{\phi}}\mu _{{r}}}{M_{{p}} \sqrt{t}}}}}C_{{h}}{m}^{3}{A_{{\phi}}}^{2}{\mu _{{r}}}^{2}\nonumber
\ee

In order to obtain the integration constants $C_h$ and $A_{\phi}$ we match the energy densities above, evaluated  at the transition time $t_{1A}$ with the corresponding values from the deSitter phase, given in (\ref{dSx}) to (\ref{dSrad}). As before, $t_{1A}$ is given by (\ref{ttrans}). For $A_{\phi}$  this leads to:
\be
A_{{\phi}}=\frac{2}{3}\,{\frac {\mu _{{p}}{M_{{p}}}^{2}{3}^{3/4}{\xi }^{3/4}}{{\eta }^{5/4}{m}^{3/2}}}\label{aphi}
\ee
The expression for $C_h$ for general $\mu_p$ turns out to be messy, but it can be simplified in the limit $\mu_p\gg 1$ to:
\be
C_{{h}}=-\frac{1}{3}\,{\frac { \sqrt{3}\mu _{{p}} \left( -4\,{\mu _{{r}}}^{2}+4\,\mu _{{p}}\mu _{{r}}+\mu _{{p}} \right)
{{\rm e}^{-4\,\mu _{{r}}}}}{{M_{{p}}}^{2}{\mu _{{r}}}^{2}}}\label{Ch}
\ee

Above we have used $\xi\sim \frac{3\sqrt{3}}{\mu_p}$ and $\eta\sim 4\sqrt{3}\mu_p$, expressions valid in the $\mu_p\gg 1$ limit.

Since we have $\mu_r\neq 0$ it we expect a minimum of the ratio $\rs$ to develop. Re-expressing  $\dot{\rs}=0$ by use of the definition of $\rs$, the ratio between the hidden sector and the radiation energy densities, and of the equations (\ref{hidradhm}) and (\ref{radradhm}) we get,
\be
\Ga\left(\rho_h+\rho_r\right)=2\rho_r\dot{\phi}\frac{\mu_r}{M_p}\label{critcondradhm}
\ee

Defining $t_c$ the time at which the above equation holds, we find that
\be
\rsm\sim \frac{2m^3\dot{\phi}(t_c)\mu_r}{M_p\rho_{\ga}(t_c)}=\frac{8}{3}{\frac {{m}^{3}\mu_{{r}}A_{{\phi}}\sqrt {t_{{c}}}
}{{M_{{p}}}^{3}}}\label{rminradhm}
\ee
From (\ref{critcondradhm}) we can also obtain $t_c$ by using the solutions we have for the energy densities in (\ref{radradtrans}) and (\ref{hradtrans}) . In order to get a closed form we will need to do some approximation of the exponent in the denominator of (\ref{hradtrans}). Since $t>t_{1A}$ we can show using $t_{1A}$ from (\ref{ttrans}) and $A_{\phi}$ from (\ref{aphi}) that $\frac{4 A_{\phi}\mu_r}{M_p\sqrt{t}}\leq 4\mu_r$. Since $\mu_r\ll 1$ we will truncate the expansion of the exponential at terms of  $\cO(\mu_{r}^2)$, leading to:
\be
t_{{c}}=1024\,{\frac {{C_{{h}}}^{2}{m}^{6}{A_{{\phi}}}^{6}{\mu _{{r}}}^{6}}{{M_{{p}}}^{2} \left( {M_{{p}}}^{2}+8\,C_{{h}}{m}^{3}{A_{{\phi}}}^{2}{\mu _{{r}}}^{2} \right) ^{2}}} \label{tcritradtrans}
\ee
Going back to (\ref{rminradhm}) we get, after some algebra, the following form for $\rsm$:
\be
\rsm={\frac{4}{\sqrt{3}}}\, {\frac{{\mu _{{r}}}^{2}}{{\mu _{{p}}}^{2}}}{{\rm e}^{-4\,\mu _{{r}}}} \left| \frac{1}{{\rm e}^{-4\mu_r}-\mu_p\left(\mu_p+4\mu_r(\mu_p-\mu_r)\right)^{-1}} \right| \nonumber
\ee
We can double expand this expression in $\mu_r$ and $\mu_p^{-1}$ and keep only the leading terms:
\be
\rsm\sim \frac{1}{18}\frac{\sqrt{3}(3+8\mu_r)}{\mu_p^2}+\cO(\mu_r^2,\mu_p^{-3})\label{rsmfinal}
\ee

In the limit $\mu_r\to 0$ we get the result in eq.(\ref{rfinalrad0}).

\end{document}